# Deep Learning for the Degraded Broadcast Channel

Erik Stauffer, Andy Wang, and Nihar Jindal

**Abstract:**
Machine learning has shown promising results for communications system problems. We present results on the use of deep auto-encoders in order to learn a transceiver for the multiuser degraded broadcast channel, and see that the auto encoder is able to learn to communicate on this channel using superposition coding. Additionally, the deep neural net is able to determine a bit labeling and optimize the per user power allocation that depends on the per user SNR.

**Introduction:**
Communication and information theory have been rich areas of research. Wireless, for example, has been a major success story with results going from research to practice, most recently with the cellular smartphone revolution. More recently, deep learning has revolutionized a number of problems such as image and speech recognition.

Although at first glance there may not seem to be much of a connection between communication/information theory with machine learning, it turns out they are deeply tied together. For example, the image detection problem is actually a channel coding problem. Some transmitter selects a picture of a cat or dog. This image is then interfered with, distorted, and noise is added. The detector, typically implemented as a convolutional neural network, plays the role of the receiver that attempts to receive the message intended by the transmitter by discriminating whether the image includes a cat or a dog.

In additional to the revolutionary results in other areas, we might expect deep learning to also have some impact on the future of communication and information theory. [1] has shown some exciting results, where the authors have demonstrated that the channel coding problem can be constructed as an autoencoder. By jointly optimizing the neural network at the encoder and the decoder, the autoencoder is able to learn to communicate over a channel. On an AWGN channel, the autoencoder learns to use a QAM. On a nonlinear channel, the autoencoder learns to do a form of predistortion [2]. [3] has also shown that deep learning can also be used to "learn" to decode convolutional and polar codes. Other work that has used deep learning for communication include [11, 12, 13].

On exciting area where machine learning may have a large impact is in the realm of multiuser information theory, where numerous theoretical results exist without practical implementation. At an abstract level, neural nets are classifying machines. By iteratively adapting the Voronoi regions, they are able to approximate an arbitrary function. Perhaps deep neural nets will be able to implement the hashing functions used in many of these theoretical results.

To connect machine learning with multiuser information theory, we consider the broadcast channel [4]. In particular, we show the promise of machine learning for multiuser information theory with a simple example, where an autoencoder is able to learn to communicate over a degraded broadcast channel.

**Broadcast Channel:**
The broadcast channel is a classic multiuser information theory problem [4]. Formally, the problem consider a single cooperative transmitter that sends information to two non-cooperating receivers. The degraded broadcast channel is a special case, where one of the receivers has a channel to the transmitter that is unambiguously better than the other user. Several results exist that describe the capacity region of the degraded broadcast channel [4, 5, 6, 7, 8]. More recent results have shown the duality between the broadcast channel and the multiple access channel [9] or with common information [10].

**System Diagram:**

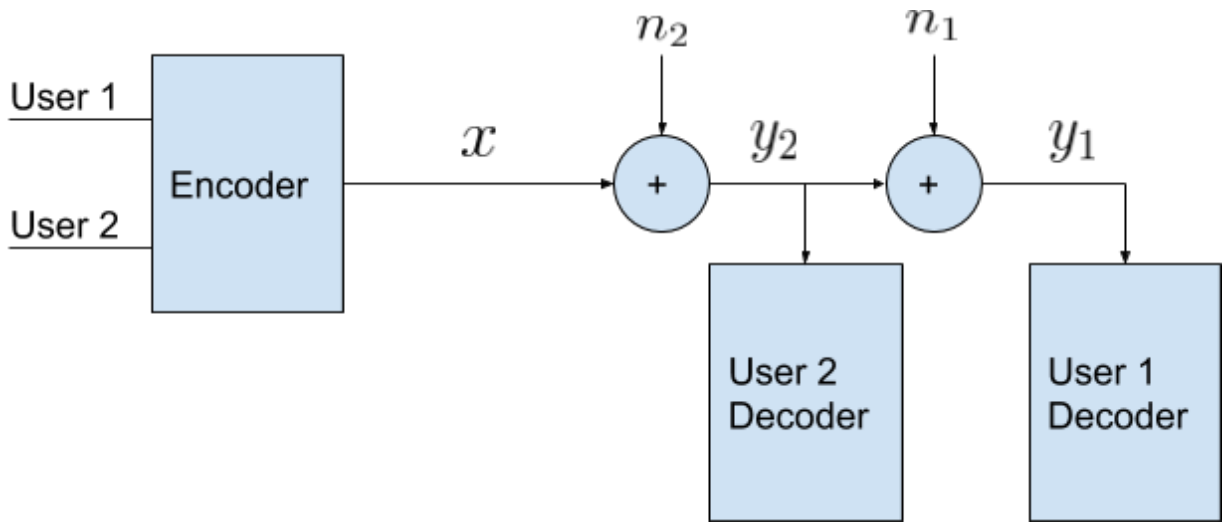

Figure 1: An AWGN degraded broadcast channel. Without loss of generality, we assume user 1 observes a degraded version of the channel compared with user 2. A joint encoder intends to transmit distinct messages to user 1 and 2. Separate decoders attempt to determine the transmitted message in the presence of noise.

The degraded broadcast channel consists of a joint transmitter, followed by two distinct decoders. The transmitter intentes to communicate one of $M_1$ messages $s_1 \in \{1, 2, ..., M_1\}$ with user 1 and one of $M_2$ messages $s_2 \in \{1, 2, ..., M_2\}$ with user 2. This corresponds to sending $k_1$ bits to user 1 and $k_2$ bits to user 2, where $M_1 = 2^{k_1}$ and $M_2 = 2^{k_2}$. The total number of bits transmitted by the encoder is $k = k_1 + k_2$, one of $M = 2^{k_1+k_2}$ messages. The transmitter selects signal $x$ subject to an average power constraint $E(||x||^2) \leq P$. Noises $n_1$ and $n_2$ are added by the channel before arriving at the two decoders, such that $y_2 = x + n_2$ and $y_1 = x + n_1 + n_2$. The decoder at user 1 and user 2 attempt to estimate messages $\hat{s}_1$ and $\hat{s}_2$.

**Simulation Configuration:**
In the following simulations, $n_1$ and $n_2$ are real valued Gaussian noise sources with variance $\sigma_2^2 = 1/SNR_2$ and $\sigma_1^2 = 1/SNR_1 - \sigma_2^2$, respectively. The transmitted signal is real valued with average power constraint $P = 1$.

**Multiuser Autoencoder Configuration:**
Following the autoencoder architecture for communication over AWGN from [1], we construct the following autoencoder structure for the degraded broadcast channel.

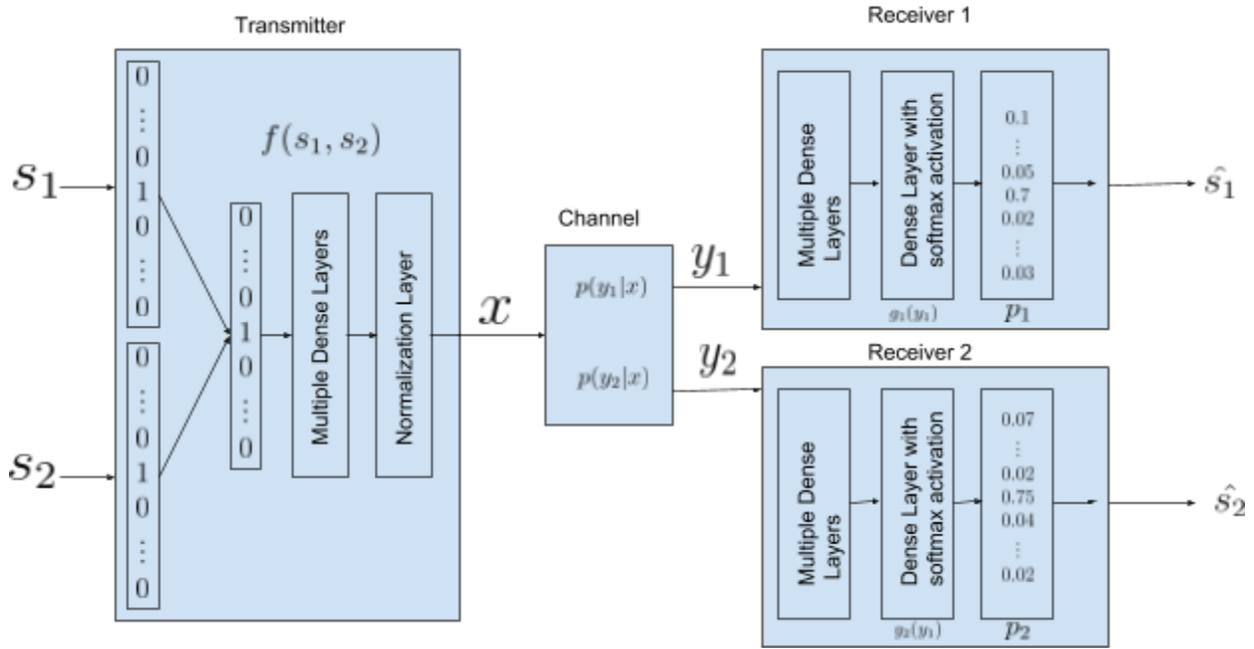

Figure 2: A broadcast communication system over a degraded AWGN channel represented as an autoencoder. The inputs $s_1$ and $s_2$ are encoded as a one-hot vector, the outputs are probability distributions over all possible message from which the most likely one with the softmax activation are picked as outputs $\hat{s}_1$ and $\hat{s}_2$.

Unless otherwise specified, the autoencoder configuration used is shown in table 1. The Adam gradient descent method was used with a learning rate of 0.001 [16], and a batch size of 1000. TensorFlow [17] was used for implementation.

| Layer | Output Dimensions (x batch_size) |
| --- | --- |
| k1 | 1 |
| k2 | 1 |
| Input (joint) | M = 2^(k1+k2) |
| Dense + ReLU | M |
| Dense + Linear | n=1 |
| Normalization | n |
| Noise | n |
| Dense + ReLU (for each decoder) | M1, M2  [M1=$2^{k_1}$, M2=$2^{k_2}$] |
| Dense + softmax (for each decoder) | M1, M2 |
| Loss: (cross entropy 1) + (cross entropy 2) | |

Table 1: Auto-encoder configuration. Two decoders are used, and the settings for both are shown.

**Results:**

By jointly optimizing the encoder and the two decoders, the autoencoder learns to communicate over the degraded broadcast channel using superposition coding. A sample of the learned constellation is shown in figure 3. User 1 is operating at 5dB SNR, and user 2 is operating at 30dB SNR. User 1 is labeled on top and user 2 is labeled on below. Note that the autoencoder has learned to give more power to user 1, with the lower SNR, vs user 2. Additionally, user 2 is labeled using a Gray coding, where the two center constellation points have identical labels.

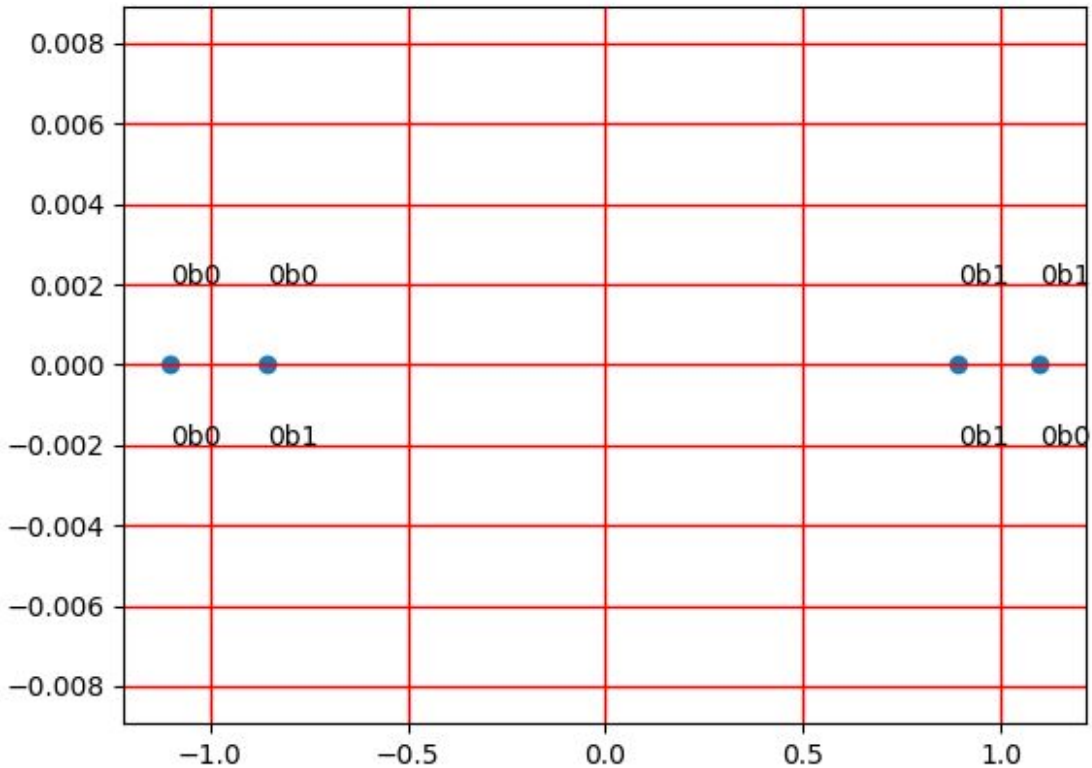

Figure 3: User 1 is labeled on top. User 2 is labeled on the bottom. User 1 information is coded by the sign of the x coordinate. User 2 information is the refinement. User 2 data is labeled using Gray coding. User 1 SNR is 5dB and User 2 SNR is 30dB.

Varying the per-user SNR changes the power allocation between the two users, as learned by the autoencoder. Figure 4 shows an example where both users have equal SNR, in this case 5dB. The power is allocated between the two users such that the spacing of the four points are equal. Exactly equal spacing between the points corresponds to a relative power weighting difference of 6dB between the users. The actual power weighting difference is 5.6 dB, presumably because the one dimensional PAMs are not capacity achieving, and additional margin is required. Again, Gray coding is found by the autoencoder to label the four points.

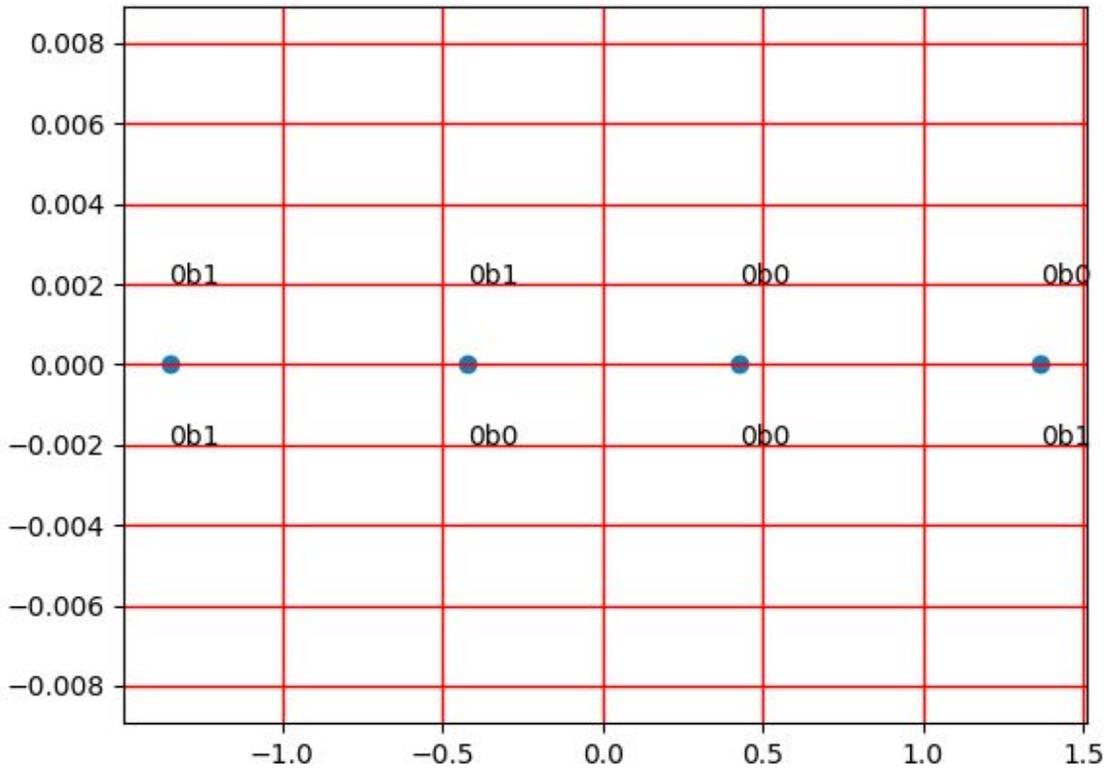

Figure 4: User 1 is labeled on top. User 2 is labeled on the bottom. User 1 and user 2 SNRs are both 5dB. Approximately equal power is assigned to the two users. Gray coding is found by the autoencoder to label the points.

Figure 5 further investigates the power allocation split between the two users vs SNR. User 1 SNR is fixed at 10dB and user 2 SNR is varied from 10dB to 30dB. Given the cross entropy loss function, the autoencoder attempts to do a form of power inversion. As the SNR for user 2 is increased, it requires less transmit power to achieve a given level of performance. Instead, more power is allocated to user 1 as the user 2 SNR increases.

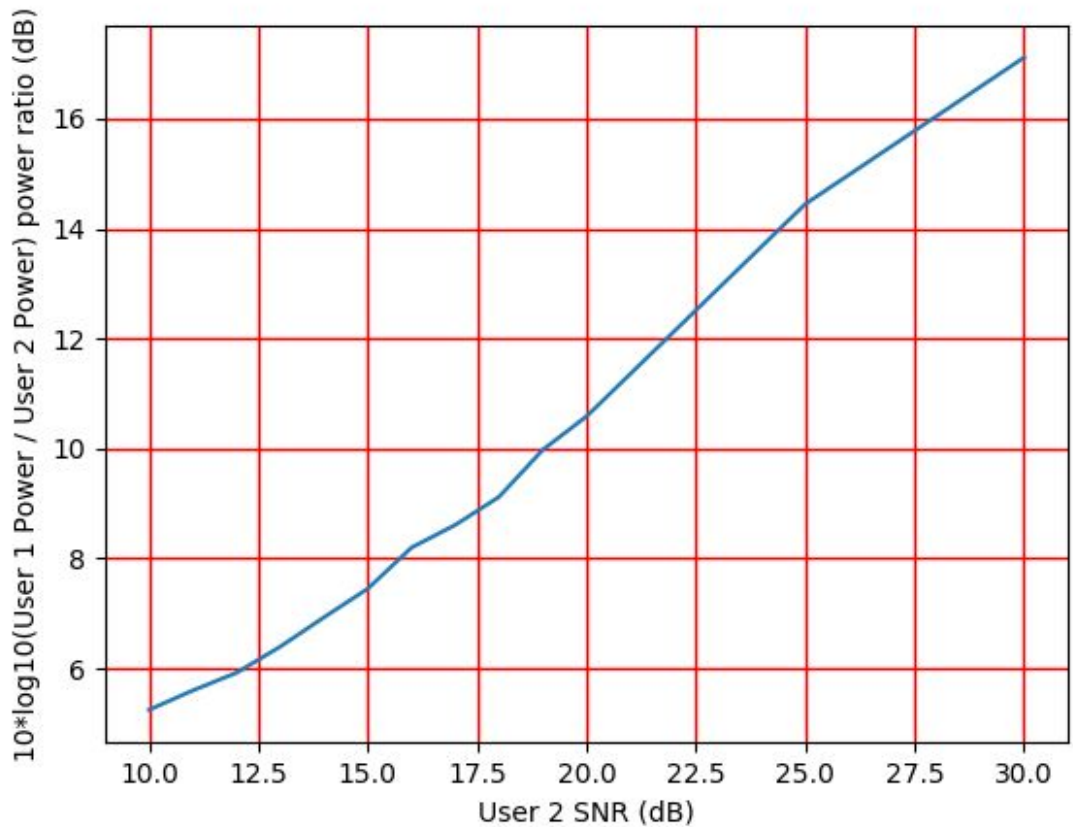

Figure 5: Power inversion. User 1 SNR is 10dB. User 2 SNR is swept from 10 to 30dB. As the SNR of user 2 is increased, move power is allocated to user 1. When the two SNRs are equal, approximately equal powers are allocated (power ratio ~=6dB). (6dB would have been optimal with capacity achieving codes)

Consistent with power inversion, at some point the user 1 SNR isn't worth supporting any data. Figure 6 shows a case where the user 1 SNR is -10dB and the user 2 SNR is 30dB. In this case, no amount of power allocated to user 1 would significantly improve the performance and the autoencoder learns to not allocate any power to this user.

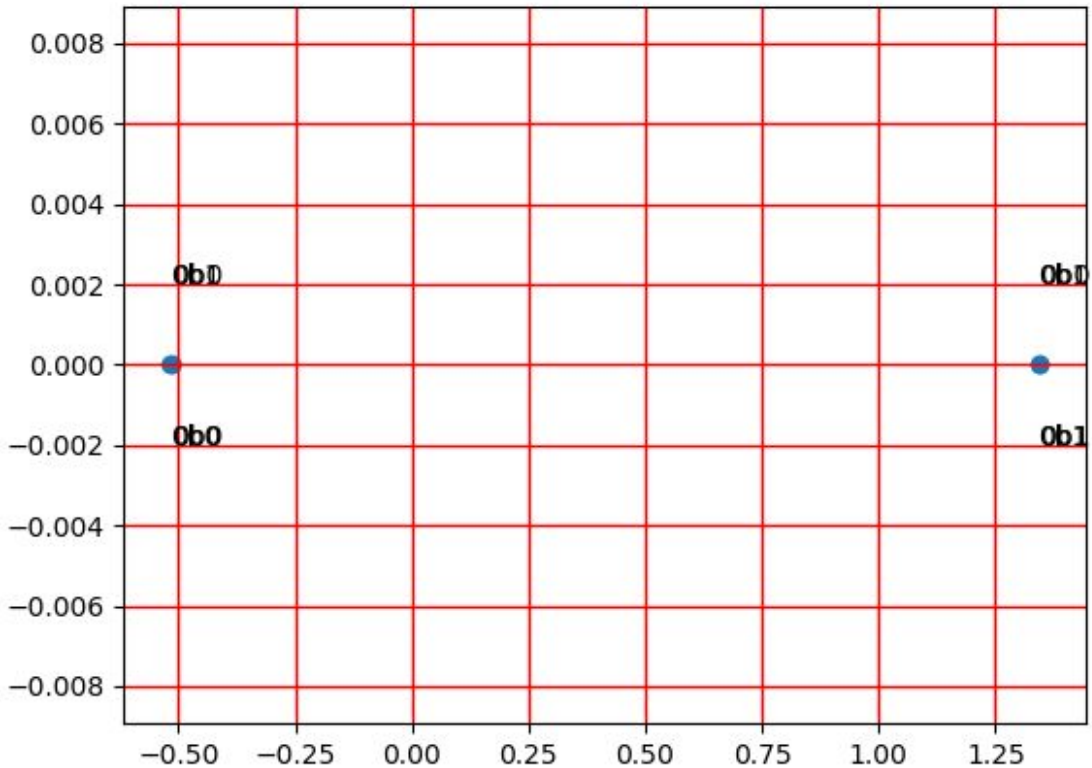

Figure 6: User 1 is labeled on top. User 2 is labeled on the bottom. User 1 SNR is -10dB and user 2 SNRs is 30dB. No power is given to user 1, and all of the power is allocated to user 2.

Table 2 shows the auto-encoder configuration for a higher order modulation example. User 2 was increased to k2=2 bits, while user 1 remained k1=1 bits. Figure 7 shows the learned constellation when the user 1 SNR is 20dB and the user 2 SNR is 40dB. The autoencoder learns to use superposition coding with a Gray coding. The user 2 constellation is a 4 PAM superimposed on top of the BPSK for user 1.

| Layer | Output Dimensions (x batch_size) |
| --- | --- |
| k1 | 1 |
| k2 | 2 |
| Input (joint) | M = 2^(k1+k2) |
| Dense + ReLU | M |
| Dense + Linear | n=1 |
| Normalization | n |
| Noise | n |
| Dense + ReLU (for each decoder) | M, M |
| Dense + ReLU (for each decoder) | M, M |
| Dense + softmax (for each decoder) | M1, M2 [M1=2^k1, M2=2^k2] |
| Loss: (cross entropy 1) + (cross entropy 2) | |

Table 2: Auto-encoder configuration for higher order modulation.

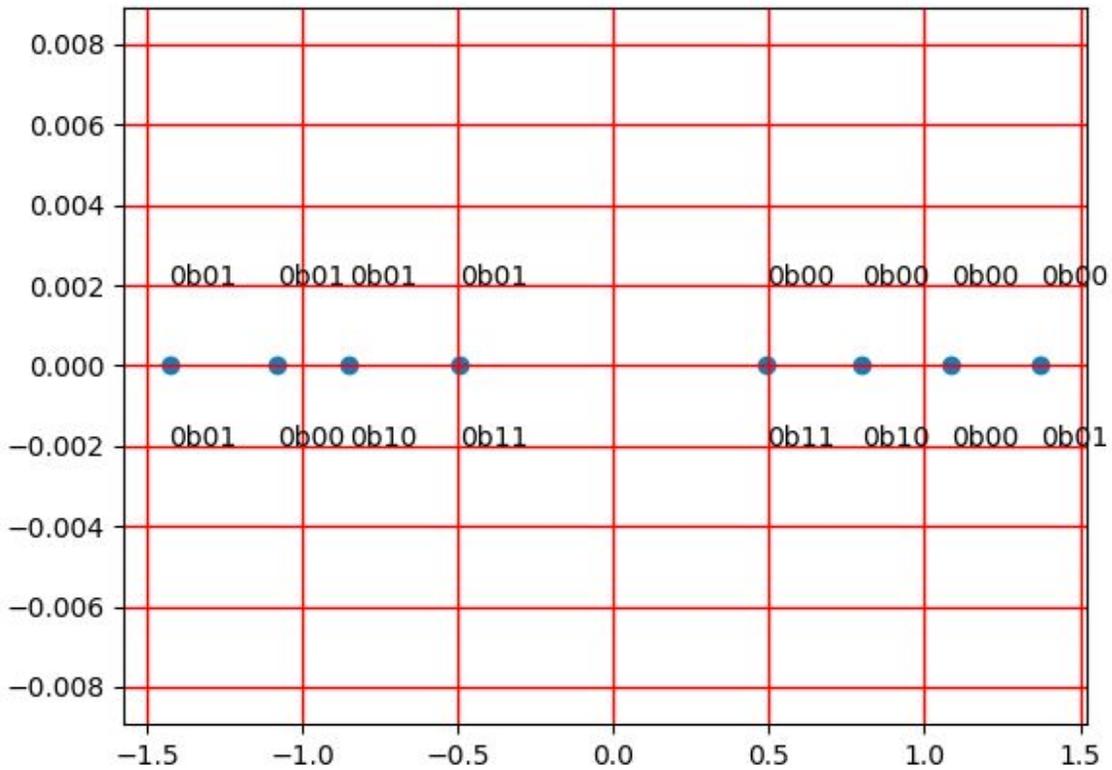

Figure 7: User 1 is labeled on top. User 2 is labeled on the bottom. User 1 SNR is 20dB and user 2 SNRs is 40dB. The autoencoder learns to do a Gray coded superposition modulation.

**Discussion:**

Since deep learning represents a heuristic based optimization, stochastic gradient descent can lead to local optima that depends on the initial conditions on non-convex problems. The bit assignment problem is non-convex, so therefore the formula presented here is also non-convex. As a result, there is some variation on the learned output from run to run.

Complexity of the one-hot encoding architecture is a concern for encoding and decoding complexity. Since the one-hot encoding vector is of size $M = 2^k$. The matrix multiply associated with the first fully connected layer is then $M * M = 2^{2k}$, and consequently complexity is exponential in the number of bits. Despite this concern, there is encouraging work by [3] and [14, 15] that gives potential ways forward for managing complexity. AlphaGo limits the otherwise exponential complexity of evaluating all future moves by using a neural network to approximate and otherwise prune the full evaluation of the future move tree. Perhaps this sort of technique can be used for communication problems as well.

**Conclusions and Future Work:**
   Deep learning has sparked an exciting time for research in a number of fields. Due to the close relationship between machine learning, one might expect an exciting future for Communication and Information theory. In this work, we present deep learning for the AWGN degraded broadcast channel and show that the autoencoder is able to learn to communicate to the two distinct users with Gray coded superposition coding.

Many future directions for this work exist. The degraded broadcast channel was presented in the context of the classic channel coding communication problem, but this communication structure also exists in image and audio processing. Another exciting extension is to treat user 2 as an interference source. By observing the signal already modulated for user 2, it may be possible to learn a form of dirty paper coding. Finally, autoencoders have shown excellent results for communicating over the non-linear channel, and an extension to the nonlinear broadcast channel may be possible.